\documentclass[12pt]{article}
\usepackage{epsfig}
\usepackage{graphicx}
\newcommand{\epem}  {\mathrm{e}^+\mathrm{e}^-}

\newcommand{\bbbar} {\mathrm{b}\bar{\mathrm{b}}}
\newcommand{\unit}[1]{\,{\mathrm{#1}}}
\begin{document}

\begin{titlepage}
\begin{center}
EUROPEAN ORGANIZATION FOR NUCLEAR RESEARCH (CERN)
\end{center}
\begin{flushright}

\bigskip
\bigskip
\bigskip

CERN-EP/2002-009\\
February 1, 2002\\
\end{flushright}

\bigskip
\bigskip
\bigskip
\bigskip
\bigskip

\begin{center}
{\LARGE \bf Search for \boldmath $\gamma\gamma \rightarrow \eta_{\mathrm b}$\\}
\vspace{4mm}
{\LARGE \bf in \boldmath $\epem$ collisions at LEP~2}
\end{center}

\bigskip
\bigskip
\bigskip

\begin{center}
{\large The ALEPH Collaboration$^{*)}$}
\end{center}

\bigskip
\bigskip
\bigskip

\begin{abstract}
\noindent
A search for the pseudoscalar meson $\eta_{\mathrm b}$ is performed 
in two-photon interactions at LEP~2 with an integrated luminosity 
of $699\unit{pb^{-1}}$ collected at $\epem$ centre-of-mass energies 
from $181\unit{GeV}$ to $209\unit{GeV}$. 
One candidate event is found in the six-charged-particle final state and
none in the four-charged-particle final state, in agreement with the total 
expected background of about one event.
Upper limits of 
\begin{center}
\begin{tabular}{lcrl}
$\Gamma_{\gamma\gamma}(\eta_{\mathrm b}) 
\times$BR($\eta_{\mathrm b} \rightarrow 4$ charged 
particles)&$<$&$48\unit{eV}$ & \\
$\Gamma_{\gamma\gamma}(\eta_{\mathrm b}) 
\times$BR($\eta_{\mathrm b} \rightarrow 6$ charged 
particles)&$<$&$132\unit{eV}$ &  
\end{tabular}
\end{center}
are obtained at 95\% confidence level, which correspond 
to upper limits of $9.0\%$ and $25\%$ on 
these branching ratios.

\bigskip
\bigskip
\bigskip

\begin{center}
{\it Submitted to Physics Letters B}
\end{center}

\bigskip

\end{abstract}

\vfill

~~~~~$^*)$ See next pages for the list of authors
\end{titlepage}

%------------------------------------------------------------------------
% authob12pt.tex
% authors' list for papers at LEP 1.5 and 2 energies
%-----------------------------------------------------------------------
\pagestyle{empty}
\newpage
\small
%
% remember the old settings
%
\newlength{\saveparskip}
\newlength{\savetextheight}
\newlength{\savetopmargin}
\newlength{\savetextwidth}
\newlength{\saveoddsidemargin}
\newlength{\savetopsep}
\setlength{\saveparskip}{\parskip}
\setlength{\savetextheight}{\textheight}
\setlength{\savetopmargin}{\topmargin}
\setlength{\savetextwidth}{\textwidth}
\setlength{\saveoddsidemargin}{\oddsidemargin}
\setlength{\savetopsep}{\topsep}
%
% text dimensions for the author list
%
\setlength{\parskip}{0.0cm}
\setlength{\textheight}{25.0cm}
\setlength{\topmargin}{-1.5cm}
\setlength{\textwidth}{16 cm}
\setlength{\oddsidemargin}{-0.0cm}
\setlength{\topsep}{1mm}
\pretolerance=10000
%%\begin{document}
%\centerline{EUROPEAN ORGANIZATION FOR NUCLEAR RESEARCH}
%\centerline{EUROPEAN LABORATORY FOR PARTICLE PHYSICS (CERN)}
%\vspace{1cm}
%\begin{flushright}CERN-EP-2000-
%\18 January 2002 - last update
%\end{flushright}
\centerline{\large\bf The ALEPH Collaboration}
\footnotesize
\vspace{0.5cm}
{\raggedbottom
\begin{sloppypar}
\samepage\noindent
A.~Heister,
S.~Schael
\nopagebreak
\begin{center}
\parbox{15.5cm}{\sl\samepage
Physikalisches Institut das RWTH-Aachen, D-52056 Aachen, Germany}
\end{center}\end{sloppypar}
\vspace{2mm}
\begin{sloppypar}
\noindent
R.~Barate,
R.~Bruneli\`ere,
I.~De~Bonis,
D.~Decamp,
C.~Goy,
S.~Jezequel,
J.-P.~Lees,
F.~Martin,
E.~Merle,
\mbox{M.-N.~Minard},
B.~Pietrzyk,
B.~Trocm\'e
\nopagebreak
\begin{center}
\parbox{15.5cm}{\sl\samepage
Laboratoire de Physique des Particules (LAPP), IN$^{2}$P$^{3}$-CNRS,
F-74019 Annecy-le-Vieux Cedex, France}
\end{center}\end{sloppypar}
\vspace{2mm}
\begin{sloppypar}
\noindent
G.~Boix,$^{25}$
S.~Bravo,
M.P.~Casado,
M.~Chmeissani,
J.M.~Crespo,
E.~Fernandez,
M.~Fernandez-Bosman,
Ll.~Garrido,$^{15}$
E.~Graug\'{e}s,
J.~Lopez,
M.~Martinez,
G.~Merino,
R.~Miquel,$^{4}$
Ll.M.~Mir,$^{4}$
A.~Pacheco,
D.~Paneque,
H.~Ruiz
\nopagebreak
\begin{center}
\parbox{15.5cm}{\sl\samepage
Institut de F\'{i}sica d'Altes Energies, Universitat Aut\`{o}noma
de Barcelona, E-08193 Bellaterra (Barcelona), Spain$^{7}$}
\end{center}\end{sloppypar}
\vspace{2mm}
\begin{sloppypar}
\noindent
A.~Colaleo,
D.~Creanza,
N.~De~Filippis,
M.~de~Palma,
G.~Iaselli,
G.~Maggi,
M.~Maggi,
S.~Nuzzo,
A.~Ranieri,
G.~Raso,$^{24}$
F.~Ruggieri,
G.~Selvaggi,
L.~Silvestris,
P.~Tempesta,
A.~Tricomi,$^{3}$
G.~Zito
\nopagebreak
\begin{center}
\parbox{15.5cm}{\sl\samepage
Dipartimento di Fisica, INFN Sezione di Bari, I-70126 Bari, Italy}
\end{center}\end{sloppypar}
\vspace{2mm}
\begin{sloppypar}
\noindent
X.~Huang,
J.~Lin,
Q. Ouyang,
T.~Wang,
Y.~Xie,
R.~Xu,
S.~Xue,
J.~Zhang,
L.~Zhang,
W.~Zhao
\nopagebreak
\begin{center}
\parbox{15.5cm}{\sl\samepage
Institute of High Energy Physics, Academia Sinica, Beijing, The People's
Republic of China$^{8}$}
\end{center}\end{sloppypar}
\vspace{2mm}
\begin{sloppypar}
\noindent
D.~Abbaneo,
P.~Azzurri,
T.~Barklow,$^{30}$
O.~Buchm\"uller,$^{30}$
M.~Cattaneo,
F.~Cerutti,
B.~Clerbaux,
H.~Drevermann,
R.W.~Forty,
M.~Frank,
F.~Gianotti,
T.C.~Greening,$^{26}$
J.B.~Hansen,
J.~Harvey,
D.E.~Hutchcroft,
P.~Janot,
B.~Jost,
M.~Kado,$^{4}$
P.~Mato,
A.~Moutoussi,
F.~Ranjard,
L.~Rolandi,
D.~Schlatter,
G.~Sguazzoni,
W.~Tejessy,
F.~Teubert,
A.~Valassi,
I.~Videau,
J.J.~Ward
\nopagebreak
\begin{center}
\parbox{15.5cm}{\sl\samepage
European Laboratory for Particle Physics (CERN), CH-1211 Geneva 23,
Switzerland}
\end{center}\end{sloppypar}
\vspace{2mm}
\begin{sloppypar}
\noindent
F.~Badaud,
S.~Dessagne,
A.~Falvard,$^{20}$
D.~Fayolle,
P.~Gay,
J.~Jousset,
B.~Michel,
S.~Monteil,
D.~Pallin,
J.M.~Pascolo,
P.~Perret
\nopagebreak
\begin{center}
\parbox{15.5cm}{\sl\samepage
Laboratoire de Physique Corpusculaire, Universit\'e Blaise Pascal,
IN$^{2}$P$^{3}$-CNRS, Clermont-Ferrand, F-63177 Aubi\`{e}re, France}
\end{center}\end{sloppypar}
\vspace{2mm}
\begin{sloppypar}
\noindent
J.D.~Hansen,
J.R.~Hansen,
P.H.~Hansen,
B.S.~Nilsson
\nopagebreak
\begin{center}
\parbox{15.5cm}{\sl\samepage
Niels Bohr Institute, 2100 Copenhagen, DK-Denmark$^{9}$}
\end{center}\end{sloppypar}
\vspace{2mm}
\begin{sloppypar}
\noindent
A.~Kyriakis,
C.~Markou,
E.~Simopoulou,
A.~Vayaki,
K.~Zachariadou
\nopagebreak
\begin{center}
\parbox{15.5cm}{\sl\samepage
Nuclear Research Center Demokritos (NRCD), GR-15310 Attiki, Greece}
\end{center}\end{sloppypar}
\vspace{2mm}
\begin{sloppypar}
\noindent
A.~Blondel,$^{12}$
\mbox{J.-C.~Brient},
F.~Machefert,
A.~Roug\'{e},
M.~Swynghedauw,
R.~Tanaka
\linebreak
H.~Videau
\nopagebreak
\begin{center}
\parbox{15.5cm}{\sl\samepage
Laboratoire de Physique Nucl\'eaire et des Hautes Energies, Ecole
Polytechnique, IN$^{2}$P$^{3}$-CNRS, \mbox{F-91128} Palaiseau Cedex, France}
\end{center}\end{sloppypar}
\vspace{2mm}
\begin{sloppypar}
\noindent
V.~Ciulli,
E.~Focardi,
G.~Parrini
\nopagebreak
\begin{center}
\parbox{15.5cm}{\sl\samepage
Dipartimento di Fisica, Universit\`a di Firenze, INFN Sezione di Firenze,
I-50125 Firenze, Italy}
\end{center}\end{sloppypar}
\vspace{2mm}
\begin{sloppypar}
\noindent
A.~Antonelli,
M.~Antonelli,
G.~Bencivenni,
F.~Bossi,
P.~Campana,
G.~Capon,
V.~Chiarella,
P.~Laurelli,
G.~Mannocchi,$^{5}$
F.~Murtas,
G.P.~Murtas,
L.~Passalacqua
\nopagebreak
\begin{center}
\parbox{15.5cm}{\sl\samepage
Laboratori Nazionali dell'INFN (LNF-INFN), I-00044 Frascati, Italy}
\end{center}\end{sloppypar}
\vspace{2mm}
%\pagebreak
\begin{sloppypar}
\noindent
A.~Halley,
J.~Kennedy,
J.G.~Lynch,
P.~Negus,
V.~O'Shea,
A.S.~Thompson
\nopagebreak
\begin{center}
\parbox{15.5cm}{\sl\samepage
Department of Physics and Astronomy, University of Glasgow, Glasgow G12
8QQ,United Kingdom$^{10}$}
\end{center}\end{sloppypar}
\vspace{2mm}
%\pagebreak
\begin{sloppypar}
\noindent
S.~Wasserbaech
\nopagebreak
\begin{center}
\parbox{15.5cm}{\sl\samepage
Department of Physics, Haverford College, Haverford, PA 19041-1392, U.S.A.}
\end{center}\end{sloppypar}
\vspace{2mm}
%\pagebreak
\begin{sloppypar}
\noindent
R.~Cavanaugh,
S.~Dhamotharan,
C.~Geweniger,
P.~Hanke,
V.~Hepp,
E.E.~Kluge,
G.~Leibenguth,
A.~Putzer,
H.~Stenzel,
K.~Tittel,
S.~Werner,$^{19}$
M.~Wunsch$^{19}$
\nopagebreak
\begin{center}
\parbox{15.5cm}{\sl\samepage
Kirchhoff-Institut f\"ur Physik, Universit\"at Heidelberg, D-69120
Heidelberg, Germany$^{16}$}
\end{center}\end{sloppypar}
\vspace{2mm}
\begin{sloppypar}
\noindent
R.~Beuselinck,
D.M.~Binnie,
W.~Cameron,
G.~Davies,
P.J.~Dornan,
M.~Girone,$^{1}$
R.D.~Hill,
N.~Marinelli,
J.~Nowell,
H.~Przysiezniak,$^{2}$
S.A.~Rutherford,
J.K.~Sedgbeer,
J.C.~Thompson,$^{14}$
R.~White
\nopagebreak
\begin{center}
\parbox{15.5cm}{\sl\samepage
Department of Physics, Imperial College, London SW7 2BZ,
United Kingdom$^{10}$}
\end{center}\end{sloppypar}
\vspace{2mm}
\begin{sloppypar}
\noindent
V.M.~Ghete,
P.~Girtler,
E.~Kneringer,
D.~Kuhn,
G.~Rudolph
\nopagebreak
\begin{center}
\parbox{15.5cm}{\sl\samepage
Institut f\"ur Experimentalphysik, Universit\"at Innsbruck, A-6020
Innsbruck, Austria$^{18}$}
\end{center}\end{sloppypar}
\vspace{2mm}
\begin{sloppypar}
\noindent
E.~Bouhova-Thacker,
C.K.~Bowdery,
D.P.~Clarke,
G.~Ellis,
A.J.~Finch,
F.~Foster,
G.~Hughes,
R.W.L.~Jones,
M.R.~Pearson,
N.A.~Robertson,
M.~Smizanska
\nopagebreak
\begin{center}
\parbox{15.5cm}{\sl\samepage
Department of Physics, University of Lancaster, Lancaster LA1 4YB,
United Kingdom$^{10}$}
\end{center}\end{sloppypar}
\vspace{2mm}
\begin{sloppypar}
\noindent
O.~van~der~Aa,
C.~Delaere,
V.~Lemaitre
\nopagebreak
\begin{center}
\parbox{15.5cm}{\sl\samepage
Institut de Physique Nucl\'eaire, D\'epartement de Physique, Universit\'e Catholique de Louvain, 1348 Louvain-la-Neuve, Belgium}
\end{center}\end{sloppypar}
\vspace{2mm}
\begin{sloppypar}
\noindent
U.~Blumenschein,
F.~H\"olldorfer,
K.~Jakobs,
F.~Kayser,
K.~Kleinknecht,
A.-S.~M\"uller,
G.~Quast,$^{6}$
B.~Renk,
H.-G.~Sander,
S.~Schmeling,
H.~Wachsmuth,
C.~Zeitnitz,
T.~Ziegler
\nopagebreak
\begin{center}
\parbox{15.5cm}{\sl\samepage
Institut f\"ur Physik, Universit\"at Mainz, D-55099 Mainz, Germany$^{16}$}
\end{center}\end{sloppypar}
\vspace{2mm}
\begin{sloppypar}
\noindent
A.~Bonissent,
P.~Coyle,
C.~Curtil,
A.~Ealet,
D.~Fouchez,
P.~Payre,
A.~Tilquin
\nopagebreak
\begin{center}
\parbox{15.5cm}{\sl\samepage
Centre de Physique des Particules de Marseille, Univ M\'editerran\'ee,
IN$^{2}$P$^{3}$-CNRS, F-13288 Marseille, France}
\end{center}\end{sloppypar}
\vspace{2mm}
\begin{sloppypar}
\noindent
F.~Ragusa
\nopagebreak
\begin{center}
\parbox{15.5cm}{\sl\samepage
Dipartimento di Fisica, Universit\`a di Milano e INFN Sezione di
Milano, I-20133 Milano, Italy.}
\end{center}\end{sloppypar}
\vspace{2mm}
\begin{sloppypar}
\noindent
A.~David,
H.~Dietl,
G.~Ganis,$^{27}$
K.~H\"uttmann,
G.~L\"utjens,
W.~M\"anner,
\mbox{H.-G.~Moser},
R.~Settles,
G.~Wolf
\nopagebreak
\begin{center}
\parbox{15.5cm}{\sl\samepage
Max-Planck-Institut f\"ur Physik, Werner-Heisenberg-Institut,
D-80805 M\"unchen, Germany\footnotemark[16]}
\end{center}\end{sloppypar}
\vspace{2mm}
\begin{sloppypar}
\noindent
J.~Boucrot,
O.~Callot,
M.~Davier,
L.~Duflot,
\mbox{J.-F.~Grivaz},
Ph.~Heusse,
A.~Jacholkowska,$^{32}$
C.~Loomis,
L.~Serin,
\mbox{J.-J.~Veillet},
J.-B.~de~Vivie~de~R\'egie,$^{28}$
C.~Yuan
\nopagebreak
\begin{center}
\parbox{15.5cm}{\sl\samepage
Laboratoire de l'Acc\'el\'erateur Lin\'eaire, Universit\'e de Paris-Sud,
IN$^{2}$P$^{3}$-CNRS, F-91898 Orsay Cedex, France}
\end{center}\end{sloppypar}
\vspace{2mm}
\begin{sloppypar}
\noindent
%\samepage
G.~Bagliesi,
T.~Boccali,
L.~Fo\`a,
A.~Giammanco,
A.~Giassi,
F.~Ligabue,
A.~Messineo,
F.~Palla,
G.~Sanguinetti,
A.~Sciab\`a,
R.~Tenchini,$^{1}$
A.~Venturi,$^{1}$
P.G.~Verdini
\samepage
\begin{center}
\parbox{15.5cm}{\sl\samepage
Dipartimento di Fisica dell'Universit\`a, INFN Sezione di Pisa,
e Scuola Normale Superiore, I-56010 Pisa, Italy}
\end{center}\end{sloppypar}
\vspace{2mm}
\begin{sloppypar}
\noindent
O.~Awunor,
G.A.~Blair,
J.~Coles,
G.~Cowan,
A.~Garcia-Bellido,
M.G.~Green,
L.T.~Jones,
T.~Medcalf,
A.~Misiejuk,
J.A.~Strong,
P.~Teixeira-Dias
\nopagebreak
\begin{center}
\parbox{15.5cm}{\sl\samepage
Department of Physics, Royal Holloway \& Bedford New College,
University of London, Egham, Surrey TW20 OEX, United Kingdom$^{10}$}
\end{center}\end{sloppypar}
\vspace{2mm}
\begin{sloppypar}
\noindent
R.W.~Clifft,
T.R.~Edgecock,
P.R.~Norton,
I.R.~Tomalin
\nopagebreak
\begin{center}
\parbox{15.5cm}{\sl\samepage
Particle Physics Dept., Rutherford Appleton Laboratory,
Chilton, Didcot, Oxon OX11 OQX, United Kingdom$^{10}$}
\end{center}\end{sloppypar}
\vspace{2mm}
%\pagebreak
\begin{sloppypar}
\noindent
\mbox{B.~Bloch-Devaux},
D.~Boumediene,
P.~Colas,
B.~Fabbro,
E.~Lan\c{c}on,
\mbox{M.-C.~Lemaire},
E.~Locci,
P.~Perez,
J.~Rander,
\mbox{J.-F.~Renardy},
A.~Rosowsky,
P.~Seager,$^{13}$
A.~Trabelsi,$^{21}$
B.~Tuchming,
B.~Vallage
\nopagebreak
\begin{center}
\parbox{15.5cm}{\sl\samepage
CEA, DAPNIA/Service de Physique des Particules,
CE-Saclay, F-91191 Gif-sur-Yvette Cedex, France$^{17}$}
\end{center}\end{sloppypar}
%\nopagebreak
\vspace{2mm}
\begin{sloppypar}
\noindent
N.~Konstantinidis,
A.M.~Litke,
G.~Taylor
\nopagebreak
\begin{center}
\parbox{15.5cm}{\sl\samepage
Institute for Particle Physics, University of California at
Santa Cruz, Santa Cruz, CA 95064, USA$^{22}$}
\end{center}\end{sloppypar}
%\pagebreak
\vspace{2mm}
\begin{sloppypar}
\noindent
C.N.~Booth,
S.~Cartwright,
F.~Combley,$^{31}$
P.N.~Hodgson,
M.~Lehto,
L.F.~Thompson
\nopagebreak
\begin{center}
\parbox{15.5cm}{\sl\samepage
Department of Physics, University of Sheffield, Sheffield S3 7RH,
United Kingdom$^{10}$}
\end{center}\end{sloppypar}
\vspace{2mm}
\begin{sloppypar}
\noindent
K.~Affholderbach,$^{23}$
A.~B\"ohrer,
S.~Brandt,
C.~Grupen,
J.~Hess,
A.~Ngac,
G.~Prange,
U.~Sieler
\nopagebreak
\begin{center}
\parbox{15.5cm}{\sl\samepage
Fachbereich Physik, Universit\"at Siegen, D-57068 Siegen, Germany$^{16}$}
\end{center}\end{sloppypar}
\vspace{2mm}
\begin{sloppypar}
\noindent
C.~Borean,
G.~Giannini
\nopagebreak
\begin{center}
\parbox{15.5cm}{\sl\samepage
Dipartimento di Fisica, Universit\`a di Trieste e INFN Sezione di Trieste,
I-34127 Trieste, Italy}
\end{center}\end{sloppypar}
\vspace{2mm}
\begin{sloppypar}
\noindent
H.~He,
J.~Putz,
J.~Rothberg
\nopagebreak
\begin{center}
\parbox{15.5cm}{\sl\samepage
Experimental Elementary Particle Physics, University of Washington, Seattle,
WA 98195 U.S.A.}
\end{center}\end{sloppypar}
\vspace{2mm}
\begin{sloppypar}
\noindent
S.R.~Armstrong,
K.~Berkelman,
K.~Cranmer,
D.P.S.~Ferguson,
Y.~Gao,$^{29}$
S.~Gonz\'{a}lez,
O.J.~Hayes,
H.~Hu,
S.~Jin,
J.~Kile,
P.A.~McNamara III,
J.~Nielsen,
Y.B.~Pan,
\mbox{J.H.~von~Wimmersperg-Toeller}, 
W.~Wiedenmann,
J.~Wu,
Sau~Lan~Wu,
X.~Wu,
G.~Zobernig
\nopagebreak
\begin{center}
\parbox{15.5cm}{\sl\samepage
Department of Physics, University of Wisconsin, Madison, WI 53706,
USA$^{11}$}
\end{center}\end{sloppypar}
\vspace{2mm}
\begin{sloppypar}
\noindent
G.~Dissertori
\nopagebreak
\begin{center}
\parbox{15.5cm}{\sl\samepage
Institute for Particle Physics, ETH H\"onggerberg, 8093 Z\"urich,
Switzerland.}
\end{center}\end{sloppypar}
}
\footnotetext[1]{Also at CERN, 1211 Geneva 23, Switzerland.}
\footnotetext[2]{Now at LAPP, 74019 Annecy-le-Vieux, France}
\footnotetext[3]{Also at Dipartimento di Fisica di Catania and INFN Sezione di
 Catania, 95129 Catania, Italy.}
\footnotetext[4]{Now at LBNL, Berkeley, CA 94720, U.S.A.}
\footnotetext[5]{Also Istituto di Cosmo-Geofisica del C.N.R., Torino,
Italy.}
\footnotetext[6]{Now at Institut f\"ur Experimentelle Kernphysik, Universit\"at Karlsruhe, 76128 Karlsruhe, Germany.}
\footnotetext[7]{Supported by CICYT, Spain.}
\footnotetext[8]{Supported by the National Science Foundation of China.}
\footnotetext[9]{Supported by the Danish Natural Science Research Council.}
\footnotetext[10]{Supported by the UK Particle Physics and Astronomy Research
Council.}
\footnotetext[11]{Supported by the US Department of Energy, grant
DE-FG0295-ER40896.}
\footnotetext[12]{Now at Departement de Physique Corpusculaire, Universit\'e de
Gen\`eve, 1211 Gen\`eve 4, Switzerland.}
\footnotetext[13]{Supported by the Commission of the European Communities,
contract ERBFMBICT982874.}
\footnotetext[14]{Also at Rutherford Appleton Laboratory, Chilton, Didcot, UK.}
\footnotetext[15]{Permanent address: Universitat de Barcelona, 08208 Barcelona,
Spain.}
\footnotetext[16]{Supported by the Bundesministerium f\"ur Bildung,
Wissenschaft, Forschung und Technologie, Germany.}
\footnotetext[17]{Supported by the Direction des Sciences de la
Mati\`ere, C.E.A.}
\footnotetext[18]{Supported by the Austrian Ministry for Science and Transport.}
\footnotetext[19]{Now at SAP AG, 69185 Walldorf, Germany}
\footnotetext[20]{Now at Groupe d' Astroparticules de Montpellier, Universit\'e de Montpellier II, 34095 Montpellier, France.}
\footnotetext[21]{Now at D\'epartement de Physique, Facult\'e des Sciences de Tunis, 1060 Le Belv\'ed\`ere, Tunisia.}
\footnotetext[22]{Supported by the US Department of Energy,
grant DE-FG03-92ER40689.}
\footnotetext[23]{Now at Skyguide, Swissair Navigation Services, Geneva, Switzerland.}
\footnotetext[24]{Also at Dipartimento di Fisica e Tecnologie Relative, Universit\`a di Palermo, Palermo, Italy.}
\footnotetext[25]{Now at McKinsey and Compagny, Avenue Louis Casal 18, 1203 Geneva, Switzerland.}
\footnotetext[26]{Now at Honeywell, Phoenix AZ, U.S.A.}
\footnotetext[27]{Now at INFN Sezione di Roma II, Dipartimento di Fisica, Universit\`a di Roma Tor Vergata, 00133 Roma, Italy.}
\footnotetext[28]{Now at Centre de Physique des Particules de Marseille, Univ M\'editerran\'ee, F-13288 Marseille, France.}
\footnotetext[29]{Also at Department of Physics, Tsinghua University, Beijing, The People's Republic of China.}
\footnotetext[30]{Now at SLAC, Stanford, CA 94309, U.S.A.}
\footnotetext[31]{Deceased.}
\footnotetext[32]{Also at Groupe d' Astroparticules de Montpellier, Universit\'e de Montpellier II, 34095 Montpellier, France.} 
\setlength{\parskip}{\saveparskip}
\setlength{\textheight}{\savetextheight}
\setlength{\topmargin}{\savetopmargin}
\setlength{\textwidth}{\savetextwidth}
\setlength{\oddsidemargin}{\saveoddsidemargin}
\setlength{\topsep}{\savetopsep}
\normalsize
\newpage
\pagestyle{plain}
\setcounter{page}{1}

\newpage

\pagenumbering{arabic}
\setcounter{page}{1}

\section{Introduction}

The $\bbbar$ ground state, the $\eta_{\mathrm b}$ meson, has not yet been 
observed. Because of their initial state, two-photon 
collisions are well suited for the study of pseudoscalar mesons, for which 
$J^{PC} = 0^{-+}$. 
The high $\gamma\gamma$ cross section and the high LEP luminosity and 
energy, as well as the low background 
from other processes, make LEP~2 a good environment to search for 
this meson. 

Theoretical estimates (from pertubative QCD and lattice nonrelativistic 
QCD) of the mass difference, $\Delta m$, between the 
$\eta_{\rm b}$ and the $\Upsilon$ 
($m_\Upsilon = 9.46$\,GeV/$c^2$) are summarized in Table~\ref{tab-mass} 
and those of the partial decay width of the $\eta_{\rm b}$ into two photons, 
$\Gamma_{\gamma\gamma}(\eta_{\mathrm b})$, in Table~\ref{tab-width}. For 
the former, values ranging from $\Delta m = 34\unit{MeV}/c^2$ to 
$141\unit{MeV}/c^2$ are obtained. For the latter, a 
value of $\Gamma_{\gamma\gamma}(\eta_{\mathrm b}) = 
557\pm 85$\,eV, chosen in this letter, is obtained from the 
average of the first order estimates (488\,eV) shifted by 69\,eV at the 
second order in $\alpha_s$. It yields an exclusive $\eta_{\rm b}$ production 
cross section of $0.304 \pm 0.046$\,pb in ${\rm e}^+{\rm e}^-$ collisions at 
$\sqrt{s}=197$\,GeV. The branching ratios of the $\eta_{\rm b}$ into four and 
six charged particles are estimated as in Ref.~\cite{etabascona} 
to be 2.7\% and 3.3\% 
respectively. (The same estimate gives 9.9\% for the $\eta_{\rm c}$ decay 
branching fraction into four charged particles, in agreement with the 
measured value of $9.3\pm1.8$\%~\cite{RPP}.) 
Six and seven exclusive $\eta_{\rm b}$ 
are therefore expected to be produced in the 699\,${\rm pb}^{-1}$ of data 
collected by ALEPH above the WW threshold, in the four- and 
six-charged-particle final states, respectively. 

A measurement of the $\eta_{\mathrm b}$ mass and of its decay modes would
therefore provide a test of pQCD and NRQCD~\cite{balireview,barnesgg,godfrey}. 
Searches have already been
conducted by the CUSB and CLEO Collaborations 
in the cascade decay of the $\Upsilon$(3S): 
%via the singlet $P$ state ${\mathrm h}_{\mathrm b}$: 
the CUSB Collaboration finds for the product of the branching ratios 
${\mathrm {BR}}(\Upsilon(3{\mathrm S}) \rightarrow \pi \pi 
{\mathrm h}_{\mathrm b}) \times
{\mathrm {BR}}({\mathrm h}_{\mathrm b}\rightarrow \gamma \eta_{\mathrm b}) 
< 0.45\%$ at 90\% C.L.\ for an $\Upsilon$-$\eta_{\mathrm b}$ splitting between 
$50\unit{MeV}/c^2$ and $110\unit{MeV}/c^2$~\cite{cusbetab}. The CLEO 
Collaboration has published a 
90\% C.L.\ upper limit on the product of the branching ratios 
${\mathrm {BR}}(\Upsilon(3{\mathrm S}) \rightarrow \pi^+ \pi^-
{\mathrm h}_{\mathrm b}) \times 
{\mathrm {BR}}({\mathrm h}_{\mathrm b}\rightarrow \gamma \eta_{\mathrm b})$ 
of about 0.1\% for the $\eta_{\mathrm b}$ mass range from $9.32\unit{GeV}/c^2$ 
to $9.46\unit{GeV}/c^2$ with a photon energy ranging 
from $434\unit{MeV}$ to 
$466\unit{MeV}$ and the ${\mathrm h}_{\mathrm b}$ mass restricted to 
$9.900 \pm 0.003\unit{GeV}/c^2$~\cite{cleoetab}.

In this letter, a search is presented for the $\eta_{\mathrm b}$ 
meson via its decay into four and six charged particles. The search is 
performed in quasi-real two-photon interactions where the meson 
is produced exclusively. This letter is organized as follows. 
A description of the ALEPH detector is given in Section 2. 
The data analysis with event selection, efficiency 
calculation, background estimate and systematic 
uncertainty determination is described in Section 3.
The results of the search are presented in 
Section 4. Finally, in Section 5 a summary is given.

\begin{table}[t]
\begin{center}
\vspace{3mm}
\caption{\small Estimates for the mass splitting 
$\Delta m = m(\Upsilon) - m(\eta_{\mathrm b})$ from QCD calculations. 
\label{tab-mass}}
\vspace{3mm}
\begin{tabular}{|lcl|}
\hline
                    & \multicolumn{1}{c}{$\Delta m$ [$\unit{MeV}/c^2$]}
                                 & Ref. \\
\hline
lattice NRQCD       & $45-100$ 
     & \cite{balireview,aida,marcantonio} \\
lattice potential   & $60-110$ 
     & \cite{balicorr} \\
pQCD                & $36-\phantom{0}55$      
                                                         & \cite{brambilla} \\
$1/m$ expansion     & $34-114$ 
                                                           & \cite{narison} \\
potential model     & $60-141$ 
                                            & \cite{barnes,eichten,ebert} \\
\hline
\end{tabular}
\end{center}
\vspace*{-0.5cm}
\end{table}

\begin{table}[t]
\begin{center}
\caption{\small Estimates for the two-photon width 
$\Gamma_{\gamma\gamma}(\eta_{\mathrm b})$. \label{tab-width}}
\vspace{3mm}
\begin{tabular}{|lcl|}
\hline
                    & \multicolumn{1}{c}{$\Gamma_{\gamma\gamma}(\eta_{\mathrm b})$} [$\unit{eV}$]  & Ref. \\
\hline
\multicolumn{3}{|c|}{estimates $\cal{O}(\alpha_{\mathrm s})$} \\
\hline
potential model     & $500~\pm~30$    & \cite{fabiano}   \\
potential model, $\Gamma_{\epem}(\Upsilon)$ & 
$490~\pm~40$    & \cite{fabiano}   \\
NRQCD               & $460\phantom{~\pm~00}$ & \cite{schuler}   \\
NRQCD, $\Gamma_{\epem}(\Upsilon)$ & 
$501\phantom{~\pm~00}$   & \cite{melnikov}   \\
\hline
\multicolumn{3}{|c|}{estimates ${\cal{O}}(\alpha_{\mathrm s}^2)$} \\
\hline
NRQCD, $\Gamma_{\epem}(\Upsilon)$ & 
$570~\pm~50$ & \cite{melnikov}   \\
\hline
\hline
used in this letter & $557~\pm~85$ & \\
\hline
\end{tabular}
\end{center}
\vspace*{-0.5cm}
\end{table}

\section{ALEPH Detector}

A detailed description of the ALEPH detector and its performance 
can be found in
Ref.~\cite{performance}. The central part of the ALEPH detector is 
dedicated to the reconstruction of the trajectories of charged
particles. The trajectory of a charged particle emerging from the 
interaction point is measured by a two-layer silicon strip 
vertex detector (VDET), a cylindrical drift chamber (ITC) and a large time 
projection chamber (TPC). The three tracking detectors are immersed in a 
$1.5\unit{T}$ axial magnetic field provided by a superconducting solenoidal 
coil. Together they measure charged particle transverse 
momenta with a resolution 
of $\delta p_{\mathrm t}/p_{\mathrm t} = 6 \times 10^{-4} p_{\mathrm t} 
\oplus 0.005$ ($p_{\mathrm t}$ in GeV/$c$). The TPC also provides a 
measurement of the specific ionization 
${\mathrm d}E/{\mathrm d}x_{\mathrm{meas}}$. An estimator may be formed 
to test a particle hypothesis, $\chi_{\mathrm h} = 
({\mathrm d}E/{\mathrm d}x_{\mathrm{meas}}-
{\mathrm d}E/{\mathrm d}x_{\mathrm{exp,h}}) / \sigma_{\mathrm{exp,h}}$, where 
${\mathrm d}E/{\mathrm d}x_{\mathrm{exp,h}}$ and 
$\sigma_{\mathrm{exp,h}}$ denote the expected specific ionization and 
the estimated uncertainty for the particle hypothesis h, respectively.

Photons are identified in the electromagnetic calorimeter 
(ECAL), situated between the TPC and the coil. The ECAL 
is a lead/proportional-tube 
sampling calorimeter segmented in $0.9^{\circ} \times 0.9^{\circ}$ projective 
towers read out in three sections in depth. It has a total thickness 
of 22 radiation lengths and yields a relative energy resolution of 
$0.18/\sqrt{E} + 0.009$, with $E$ in GeV, for isolated photons. Electrons are 
identified by their transverse and longitudinal shower profiles in 
ECAL and their specific ionization in the TPC.

The iron return yoke is instrumented with 23 layers of streamer tubes and 
forms the hadron calorimeter (HCAL). The latter provides a relative 
energy resolution of charged and neutral hadrons of $0.85/\sqrt{E}$, 
with $E$ in GeV. Muons are distinguished from hadrons by their characteristic 
pattern in HCAL and by the muon chambers, composed of two double-layers 
of streamer tubes outside HCAL. 

The information from the tracking detectors and the calorimeters are 
combined in an energy-flow algorithm~\cite{performance}. For each event, the 
algorithm provides a set of charged and neutral reconstructed particles, 
called {\em energy-flow objects} in the following.

\section{Analysis}

\subsection{Event Selection}

The search is performed in the four- and six-charged-particle 
modes, where four (or six) 
charged energy-flow objects with a net charge zero are required. 
In order to keep the efficiency high, loose selection cuts are chosen.
No attempt is made to reconstruct K$_{\mathrm S}$ mesons at this stage. 
The ${\mathrm d}E/{\mathrm d}x$ measurement, when available, 
must be consistent with the pion or kaon hypothesis 
($\chi^2_{\mathrm {\pi,K}} < 9$); the more likely hypothesis 
is used for mass assignment. When no ${\mathrm d}E/{\mathrm d}x$ information 
is available the pion mass is assigned to the particle. No neutral 
energy-flow object with $E > 1 \unit{GeV}$ must be present
within $20^{\circ}$ of the beam axis. No muon and no electron (as defined 
by the ECAL) 
must be observed. Events are also excluded if a photon conversion is 
detected, where the electron and positron are identified by requiring 
$\chi^2_{\mathrm e} < 9$, and the pair invariant mass 
is smaller than $25\unit{MeV}/c^2$.

The total transverse momentum of charged particles in 
the event ($\sum \vec{p}_{t,i}$) must be smaller than 
$250 \unit{MeV}/c$. The energy-flow objects in the event 
are boosted into their centre-of-mass frame and 
the thrust is computed in this frame. 
The thrust axis must form an angle $\theta_{\mathrm{thrust}}$ 
larger than $45^{\circ}$ 
with respect to the beam axis to reject events from the 
$\gamma\gamma$ continuum background. The 
$\gamma\gamma \rightarrow \tau^+\tau^-$ background is 
reduced to a negligible fraction by the rejection of events in which 
both hemispheres, as defined by the thrust axis, have a 
net charge of $\pm$1 and an invariant mass less than $1.8\unit{GeV}/c^2$. 

\subsection{Signal Efficiency}

Selection and reconstruction efficiencies are studied with events 
generated with PHOT02 \cite{alex} in which the $\eta_{\mathrm b}$ 
mass is set to $9.4\unit{GeV}/c^2$ and the total width to 
$7\unit{MeV}/c^2$. The width is calculated under the assumption 
that the two-gluon 
decay is dominant~\cite{RPP,close,cleo}. Four samples of 2500 events each 
are generated for the final state with four charged particles 
(2($\pi^+\pi^-$), $\pi^+\pi^-$K$^+$K$^-$, 2(K$^+$K$^-$), 
K$_{\mathrm S}$K$^+ \pi^-$). Four other samples of 2500 events each 
are generated for the final state with six charged particles 
(3($\pi^+\pi^-$), 2($\pi^+\pi^-$)K$^+$K$^-$, 
$\pi^+\pi^-$2(K$^+$K$^-$), 3(K$^+$K$^-$)).
For the decays, it is assumed that the momenta are distributed 
according to phase space. 
The event samples are passed through the detector simulation and 
reconstruction programs. 
The mass resolution of the selected events is about 
$0.14\unit{GeV}/c^2$ and is dominated by wrong mass assignment from 
$\pi$-K misidentification. A signal region between 
$9.0\unit{GeV}/c^2$ and $9.8\unit{GeV}/c^2$ is chosen.
The event selection efficiencies averaged over the four 
decay channels are found to be 16.7\% and 9.3\% 
for the four- and six-charged-track channels, respectively. 

\subsection{Systematic Uncertainties}

The lack of knowledge of the decay modes and kinematics of the 
$\eta_{\mathrm b}$ meson is the source of the dominant systematic 
uncertainties in the analysis. 
The uncertainty on the selection efficiency due to the unknown decay mode 
of the $\eta_{\mathrm b}$ meson is estimated from the spread of the 
efficiencies of the four simulated decay modes. 
The relative uncertainty is 7.5\% and 20.4\% for the four- and 
six-charged-particle final states. 
In order to check the effect of the selection efficiency due to the 
assumption of phase space decays, the $\eta_{\mathrm b}$ is forced to 
decay into a pair of $\phi$ mesons, each giving two charged kaons. 
In this case a relative increase of 10\% in the detection efficiency 
is found. 

Further studies are performed without the final cut on neutral 
energy or with modified cuts on $\sum \vec{p}_{t,i}$, 
$\theta_{\mathrm{thrust}}$, and hemisphere mass. 
An uncertainty of $5.5\%$ is estimated.
The limited statistics of simulated events contribute an uncertainty of 
$2.4\%$ and $3.2\%$ for the two decay modes, respectively.

A total relative uncertainty of 9.7\% (21.4\%) on the selection efficiency is 
calculated for the four- (six-) charged-track channel.

\subsection{Background Estimate}

The background estimate suffers from the low statistics of the simulated 
events selected and from the poor description of the shape of the 
invariant mass spectra. The background, dominated by 
$\gamma\gamma$ continuum production, is therefore estimated from data 
by means of a fit to the ratio of the mass spectra after all cuts are applied 
and before the final cuts on $\sum \vec{p}_{t,i}$, 
$\theta_{\mathrm{thrust}}$, and hemisphere mass are applied. 
The ratio is fitted with an exponential function 
up to $m=6\unit{GeV}/c^2$ ($m=7\unit{GeV}/c^2$) for the 
four- (six-) charged-particle topology. The 
average of the values of this function 
at $m=6\unit{GeV}/c^2$ ($m=7\unit{GeV}/c^2$) and at 
$m=9.4\unit{GeV}/c^2$ is 
then multiplied by the number of events in the signal 
region before the final cuts to obtain the background estimate. 
Half of the difference between these two values is taken 
as the systematic uncertainty on the estimate.
The background in the signal region is determined to be 
$0.30 \pm 0.25$ ($0.70 \pm 0.34$) events for 
the four- (six-) charged-particle topology.
%the sum is $1.00 \pm 0.47$.

\section{Results}

\begin{figure}
\begin{center}
{\epsfig{file=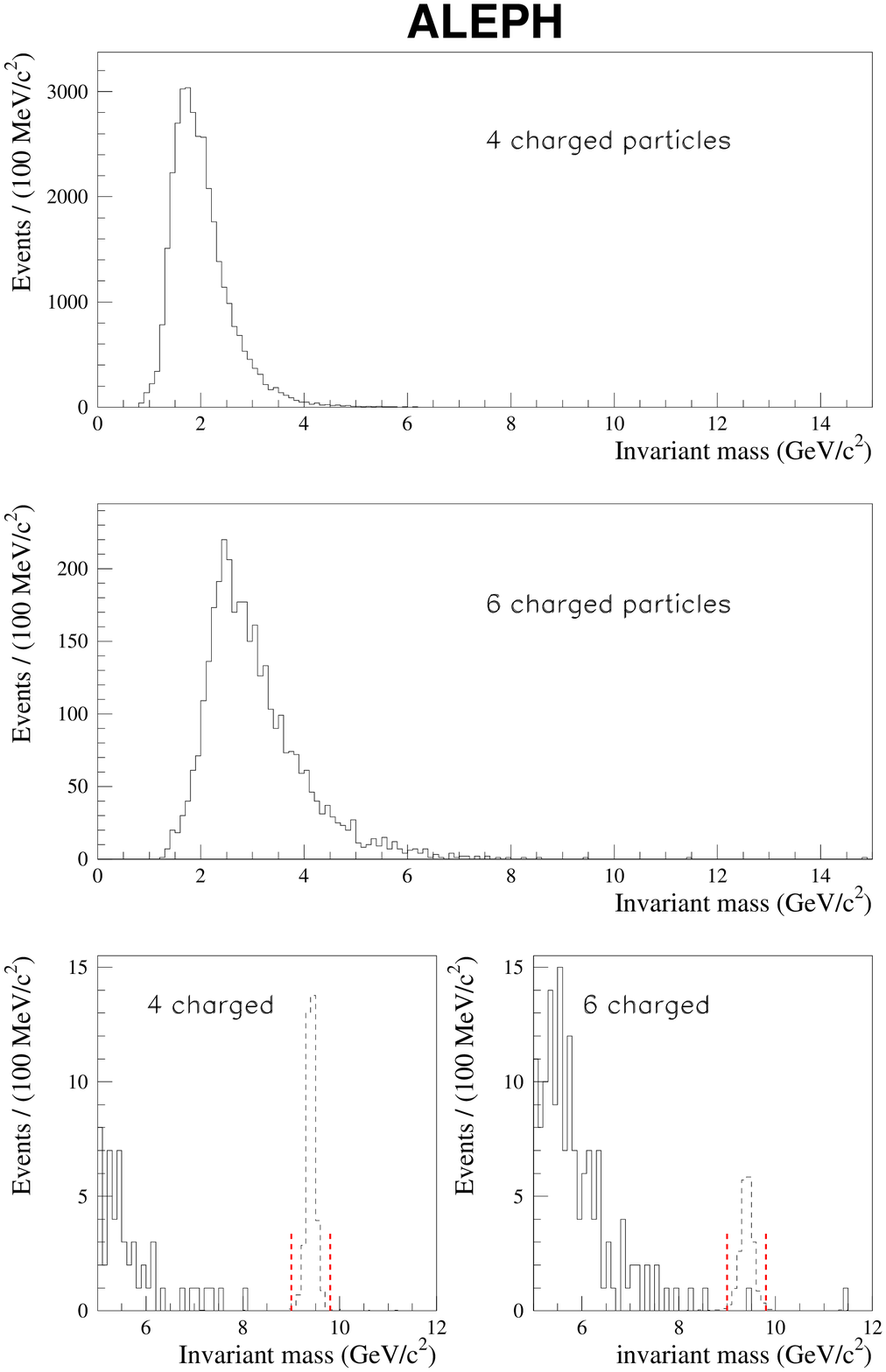,width=13.5cm}}
\caption{\small Invariant mass distribution of selected events for
four- and six-charged-particle final states (solid line: data). The dashed 
line represents the expected signal for a 100\% branching ratio into the 
mode under consideration. The signal region is indicated 
by the vertical dashed lines.
\label{allsigmass}}
\end{center}
\end{figure}

Invariant mass spectra of the selected events  
are shown in Fig.~\ref{allsigmass}. 
A total of 33727 (3432) events is selected in the
four- (six-) charged-particle final states. In the signal region, 
only one event is found in the six-prong topology. 

\subsection{Cross Section Upper Limit}

From the knowledge of the background $b$ and the efficiency 
$\varepsilon$, the observed number 
of events $n$ is converted~\cite{zech} into an upper limit on the 
signal events $\mu$ into a confidence level $\alpha$ given by 
\begin{displaymath}
1 - \alpha = 
    \frac{\int g(b) f(\varepsilon) \sum_{i=0}^n P(i \mid \mu \varepsilon + b) 
                                {\mathrm d}\varepsilon {\mathrm d}b} 
         {\int g(b) \sum_{i=0}^n P(i \mid b) 
                                {\mathrm d}b} \ \ ,
%\nonumber
\end{displaymath}
where $P(j \mid x)$ is the Poisson probability that $j$ events be 
observed, when $x$ are expected. 
The probability density functions for the background $g(b)$ 
and the efficiency $f(\varepsilon)$ are 
assumed to be Gaussian, but restricted to the range where 
$b$ and $\varepsilon$ are positive. Upper limits 
of 3.06 (4.69) events at 95\% confidence level are calculated for the 
four- (six-) prong topology. 
This translates into the upper limits 
\begin{center}
\begin{tabular}{lcrl}
$\Gamma_{\gamma\gamma}(\eta_{\mathrm b}) 
\times$BR($\eta_{\mathrm b} \rightarrow 4$ charged 
particles)&$<$&$48\unit{eV}$ & \\
$\Gamma_{\gamma\gamma}(\eta_{\mathrm b}) 
\times$BR($\eta_{\mathrm b} \rightarrow 6$ charged 
particles)&$<$&$132\unit{eV}$ & .
\end{tabular}
\end{center}

With a two-photon width of $557\pm 85$\,eV,  upper limits on the 
branching ratios 
BR($\eta_{\mathrm b} \rightarrow 4$ charged particles) $<$ 9.0\% and 
BR($\eta_{\mathrm b} \rightarrow 6$ charged particles) $<$ 25\% are 
derived.

\subsection{Mass of the Candidate}

\begin{figure}
\begin{center}
{\epsfig{file=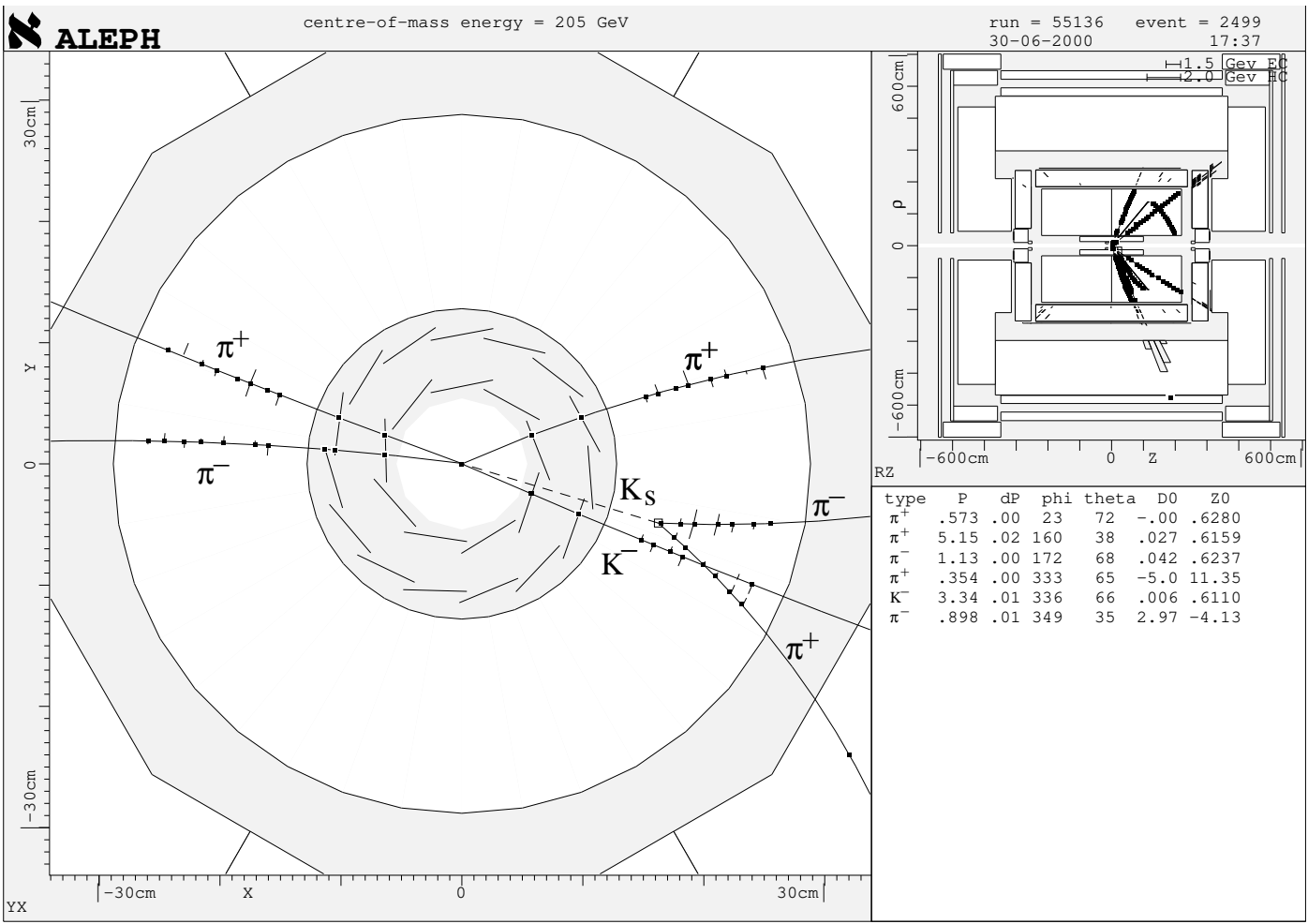,width=13.0cm,clip}}
\caption{\small An $r\phi$ view of the 
$\eta_{\mathrm b} \rightarrow {\mathrm K}_{\mathrm S}$K$^- \pi^+ \pi^- \pi^+$ 
candidate event with the reconstructed mass of 
$9.30 \pm 0.02 \pm 0.02\unit{GeV}/c^2$, 
selected in the signal region. The track coordinates recorded 
in the VDET and the ITC are shown. The tracks are appropriately 
labeled. The plot to the right shows an $rz$ view of the ALEPH 
apparatus. Information is given for each track: 
particle type, momentum (GeV$/c$), momentum error (GeV$/c$), 
azimuthal and polar angle (degrees), transverse and longitudinal 
impact parameter (cm).
%Track \#~7 only 
%seen in the TPC is the second half of the curling track \#~4.
\label{etabbest-inner}}
\end{center}
\end{figure}

The raw reconstructed mass of the candidate, as obtained from the measured
momenta of the six charged particles and with
masses assigned according to the d$E$/d$x$ measurement, is 9.45\,GeV/$c^2$.
The mass estimate can be refined with additional information visible from
the event display shown in Fig.~\ref{etabbest-inner}. 
Two of the six tracks form a secondary
vertex compatible with the decay of a ${\rm K}_{\rm S}$ into $\pi^+\pi^-$.
This hypothesis is supported by the presence of a third track compatible
with a ${\rm K}^-$ ($\chi_\pi^2 = 6.0$ and $\chi_{\rm K}^2 = 3.8 \times
10^{-5}$). The secondary vertex is therefore fitted to this hypothesis, and
the five particles (three charged pions, one charged kaon and one ${\rm
K}_{\rm S}$) are forced to originate from a common primary vertex. A mass 
of $9.30\pm0.02\pm0.02\unit{GeV}/c^2$ is derived from these constraints. 

\begin{figure}
\begin{center}
{\epsfig{file=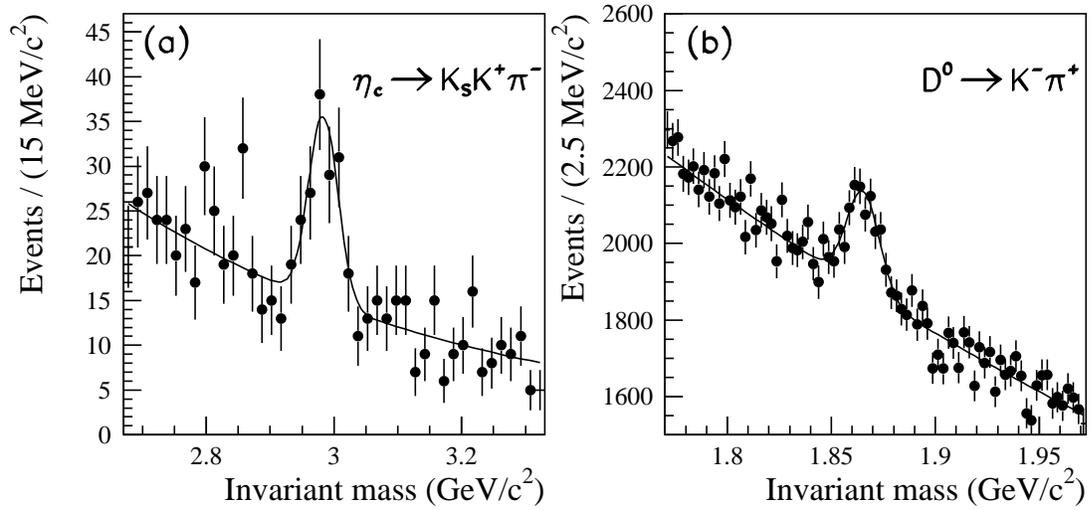,width=16.cm}}
\caption{\small (a) Invariant mass distribution of the selected events of
the K$_{\mathrm S}$K$^+ \pi^-$ control sample showing the
signal of the $\eta_{\mathrm c}$ meson. (b) The D$^0$ signal 
reconstructed in its K$^-\pi^+$ decay mode.
\label{etacandd0}}
\end{center}
\end{figure}

A control sample of $\eta_{\mathrm c}$ mesons is selected in the 
K$_{\mathrm S}$K$^+ \pi^-$ decay mode, without the final cuts 
but that on the total transverse momentum, which is relaxed to 
$\sum \vec{p}_{t,i} < 500 \unit{MeV}/c$. 
The analysis is repeated with this control sample for the study 
of the systematic uncertainty on the mass determination. 
The mass of the $\eta_{\mathrm c}$ meson is fitted and is found consistent 
with the world average value~\cite{RPP} within its statistical 
accuracy of $4.7\unit{MeV}/c^2$. A systematic uncertainty of the 
same size is assigned. The total uncertainty is then rescaled 
with the mass ratio $m$(candidate)/$m(\eta_{\mathrm c})$ and a 
systematic uncertainty of $21\unit{MeV}/c^2$ is obtained for the 
mass estimate of the $\eta_{\mathrm b}$ candidate. 
The $\eta_{\mathrm c}$ signal is 
shown in Fig.~\ref{etacandd0} together with the D$^0$ signal as observed in 
its K$^-\pi^+$ decay mode. The 
fitted D$^0$ mass agrees with the world average 
value~\cite{RPP} within its statistical accuracy of $0.9\unit{MeV}/c^2$. 
The number of observed $\eta_{\mathrm c}$ mesons is consistent with previous 
measurements~\cite{RPP,cleo,l3etac}.

\section{Summary}

With an integrated luminosity of $699\unit{pb^{-1}}$ 
collected at $\epem$ centre-of-mass energies 
between $181\unit{GeV}$ and $209\unit{GeV}$, 
the pseudoscalar meson $\eta_{\mathrm b}$ is searched for in 
its decays to four and six charged particles. 
One candidate is retained in the decay mode into six charged particles, 
while no candidate is found in the four-charged-particle decay mode. 
The candidate $\eta_{\mathrm b}$ has a reconstructed invariant 
mass of $9.30 \pm 0.02 \pm 0.02\unit{GeV}/c^2$. 
The observation of one event is consistent with 
the number of events expected from background.

Upper limits on 
$\Gamma_{\gamma\gamma}(\eta_{\mathrm b}) \times$BR of 
$48\unit{eV}$ and $132\unit{eV}$, corresponding to limits on the 
branching ratios  
BR($\eta_{\mathrm b} \rightarrow 4$ charged particles)$ < 9.0\%$ and 
BR($\eta_{\mathrm b} \rightarrow 6$ charged particles)$ < 25\%$, 
are obtained at a confidence level of 95\%.

\subsection*{Acknowledgements}

We wish to thank our colleagues in the CERN accelerator divisions for 
the successful operation of LEP. We are indebted to the engineers and 
technicians in all our institutions for their contribution to the 
excellent performance of ALEPH. Those of us from non-member 
countries thank CERN for its hospitality. We would like to thank 
Ted Barnes and Gunnar Bali for discussions.

%\newpage 
%\vfill

\end{document}